\documentclass{article}
\usepackage{spconf,amsmath,epsfig}
\usepackage{cite}

\usepackage{todonotes}

\let\oldbibliography\thebibliography
\renewcommand{\thebibliography}[1]{%
  \oldbibliography{#1}%
  \setlength{\itemsep}{0pt}%
}
\newcommand\blfootnote[1]{%
  \begingroup
  \renewcommand\thefootnote{}\footnote{#1}%
  \addtocounter{footnote}{-1}%
  \endgroup
}
\makeatletter%
\long\def\@makefntext#1{%
  \parindent 0.01em\noindent \hb@xt@ 0.1em{\hss \@makefnmark}\hskip-0.1mm\relax#1}%
\makeatother

\usepackage{nicefrac,amssymb}
\usepackage{multirow}
\usepackage{tikz}
\usepackage{caption}
\usepackage{subcaption}

\usepackage{hyperref}
\hypersetup{
    colorlinks=true,
    linkcolor=black,
    citecolor=black,
    urlcolor=blue,
}

\usepackage[overlay,absolute]{textpos}

\title{S\textsuperscript{2}-cGAN: Self-Supervised Adversarial Representation Learning for Binary Change Detection in Multispectral Images}
\name{Jose Luis Holgado Alvarez, Mahdyar Ravanbakhsh, Beg{\"u}m Demir}
\address{Faculty of Electrical Engineering and Computer Science, Technische Universit{\"a}t Berlin, Germany}
\begin{document}
%
\begin{textblock*}{8in}(5mm, 5mm)
{\scriptsize{$\copyright$ 2020 IEEE. Published in the IEEE 2020 International Geoscience \& Remote Sensing Symposium (IGARSS 2020), July 2020, Waikoloa, Hawaii, USA. Personal use of this material is permitted.}}
\end{textblock*}
\maketitle
\begin{abstract}
Deep Neural Networks have recently demonstrated promising performance in binary change detection (CD) problems in remote sensing (RS), requiring a large amount of labeled multitemporal training samples. Since collecting such data is time-consuming and costly, most of the existing methods rely on pre-trained networks on publicly available computer vision (CV) datasets. However, because of the differences in image characteristics in CV and RS, this approach limits the performance of the existing CD methods. To address this problem, we propose a self-supervised conditional Generative Adversarial Network (S\textsuperscript{2}-cGAN). The proposed S\textsuperscript{2}-cGAN is trained to generate only the distribution of unchanged samples. To this end, the proposed method consists of two main steps: 1) Generating a reconstructed version of the input image as an unchanged image 2) Learning the distribution of unchanged samples through an adversarial game. Unlike the existing GAN based methods (which only use the discriminator during the adversarial training to supervise the generator), the S\textsuperscript{2}-cGAN directly exploits the discriminator likelihood to solve the binary CD task. Experimental results show the effectiveness of the proposed S\textsuperscript{2}-cGAN when compared to the state of the art CD methods.\blfootnote{\scriptsize{Our code is available online: \href{https://gitlab.tubit.tu-berlin.de/rsim/S2-cGAN}{https://gitlab.tubit.tu-berlin.de/rsim/S2-cGAN}}}
\end{abstract}

\section{Introduction}
\label{sec:intro}

Binary change detection (CD) in multitemporal multispectral remote sensing (RS) images is a key component for monitoring environmental phenomena \cite{el2016convolutional,daudt2018fully,song2018change,saha2019unsupervised}. In the last years, Deep Neural Networks (DNNs) have achieved remarkable performance in several RS applications, including CD \cite{daudt2018fully,song2018change}. Most of the DNNs designed for CD problems require a huge amount of multitemporal labeled samples to adjust all parameters during training and reach high performance. However, collecting multitemporal labeled samples is often highly expensive and needs expertise.
To address this problem, unsupervised deep learning (DL) based approaches are recently introduced in RS to employ pre-trained networks as a generic feature extractor. El Amin et al. propose to extract deep distance images (DI) using a fully convolutional network trained for semantic segmentation in \cite{el2016convolutional}. In this work, it is shown that the extracted deep feature maps contain semantic information, which can improve the DI analysis. Similarly, Saha et al. demonstrate the effectiveness of employing semantic-aware deep feature maps to achieve a performance gain in Change Vector Analysis (CVA) \cite{saha2018unsupervised}. Bergamasco et al. propose a multilayer convolutional-autoencoder to reconstruct the input image and learn task-specific features using a deep network \cite{bergamasco2019unsupervised}. Then, multi-scale features are extracted from the pre and post change images that are analyzed for CD. In recent years, unsupervised CD methods based on deep Generative Adversarial Network (GANs) \cite{DBLP:journals/corr/IsolaZZE16} have come to the light of attention in RS. GANs are deep networks commonly used for generating realistic data (e.g., images), where the supervision is indirectly provided by an adversarial game between two independent networks: a generator ($G$) and a discriminator ($D$), which are both trained with unlabeled samples. Recently, Saha et al. adapt a GAN to solve multisensor CD, where the task is to detect changes across two images obtained by different sensors \cite{saha2019unsupervised}. In this work, the generative power of GAN is only utilized to transcode images between two domains, and an external classifier is required to detect the changes between two transcoded images. Most of the above-mentioned methods exploit DL models with proven architectures that are pre-trained on large-scale computer vision (CV) datasets (e.g., ImageNet). However, this is not a fully suitable approach in RS, because of the differences in characteristics of CV and RS images.

Unlike the aforementioned deep generative approaches, in this paper we aim at learning an adversarial representation that not only can be learned in a self-supervised fashion but also can be directly used for CD without any need for an external classifier and a pre-trained network. To this end, we propose a self-supervised conditional Generative Adversarial Network (S\textsuperscript{2}-cGAN). The S\textsuperscript{2}-cGAN is trained using a training set containing only pairs of unchanged samples that are reconstructed (generated) from a single image without using any external supervision. During the adversarial training, $G$ learns how to generate unchanged data only, whereas $D$ learns how to detect deviations from it. Such $D$ can be directly used as a change classifier, considering changes as outliers with respect to the sample distribution learned on unchanged data. According to our knowledge, the proposed S\textsuperscript{2}-cGAN is the first method that directly exploits the discriminator GAN to formulate the binary change detection problem.

\section{Proposed Approach}
\label{sec:training}
 Let $\boldsymbol X_1$ and $\boldsymbol X_2$ be two co-registered RS images acquired using a same sensor over the same geographical area at times $t_1$ and $t_2$, respectively. Both images are divided into non-overlapping $N$ patches and represented as $\boldsymbol X_1 =\{x_1^i\}_{i=1,...,N}$ and $\boldsymbol X_2 =\{x_2^i\}_{i=1,...,N}$, where $x_1^i$ and $x_2^i$ show the $i^{th}$ pair of patches associated to $t_1$ and $t_2$. In order to detect binary changes in $\boldsymbol X_1$ and $\boldsymbol X_2$, in this paper a self-supervised conditional Generative Adversarial Network (S\textsuperscript{2}-cGAN) is proposed. The proposed S\textsuperscript{2}-cGAN aims to train networks $G$ and $D$ using only pairs of patches without any land-cover change. Network $G$ learns how to generate only the unchanged pairs of patches. On the other hand, $D$ learns to distinguish pairs of unchanged samples (pixels) from those of changed samples. 
 Fig.~\ref{fig:teaser} summarize the general training strategy of the proposed S\textsuperscript{2}-cGAN. The details are provided in the following sub-sections.

 \subsection{Self-Supervised Adversarial Learning}

The proposed S\textsuperscript{2}-cGAN aims at learning a self-supervised representation from unchanged patches to estimate the likelihood of change concerning the learned distribution to detect the possible changes accordingly. In general, conditional GANs (cGANs) consist of two networks: 1) the generator network ($G$) that aims at generating realistic data, and 2) the discriminator network ($D$) that aims at discriminating real data from generated data by $G$. More specifically, the conditional GANs (cGANs) \cite{DBLP:journals/corr/IsolaZZE16} take as input an image such $I_1$ and generate a new image $\tilde{I_1}$. $D$ tries to distinguish $I_1$ from $\tilde{I_1}$ considering the conditional information given by image $I_2$, while $G$ tries to fool $D$ by producing more and more realistic images which are indistinguishable.
Isola et al.  proposed an image-to-image translation framework based on cGANs \cite{DBLP:journals/corr/IsolaZZE16}, and show that a U-Net encoder-decoder with skip connections can be used as the generator architecture together with a patch-based discriminator to transform images to different representations (conditional information). 

The proposed S\textsuperscript{2}-cGAN is inspired by \cite{DBLP:journals/corr/IsolaZZE16}, however, unlike \cite{DBLP:journals/corr/IsolaZZE16} the S\textsuperscript{2}-cGAN do not aim at realistic image translation. Instead, the S\textsuperscript{2}-cGAN exploits $G$  to learn the pattern of unchanged pair of patches. In order to train our network, we need a set of pair of multitemporal images, in which no change observed between times $t_1$ and $t_2$. Such pair can be constructed using image $\boldsymbol X_1$, and a corresponding image $\boldsymbol{\tilde{X}}_2= \{\tilde{x}_2^i\}_{i=1,...,N}$ can be defined as : $\boldsymbol{\tilde{X}}_2=\boldsymbol{X}_1+w$, where $w$ represents the noise, which is assumed to be zero-mean Gaussian for all patches in $\boldsymbol X_1$ selected from a fixed noise distribution, such that $w \sim \mathcal{N}(0,\sigma)$. Then, a set ${\cal X} = \{ (x_1^i, \tilde{x}_2^i) \}_{i=1,...,N}$ of pairs of patches selected from images $\boldsymbol{X}_1$ and $\boldsymbol{\tilde{X}}_2$ is defined. This set represents the unchanged pairs corresponding to the same geographical area and is used for learning unchanged data representation in a self-supervised fashion.
\begin{figure}[t]
\centerline{\includegraphics[width=0.85\linewidth]{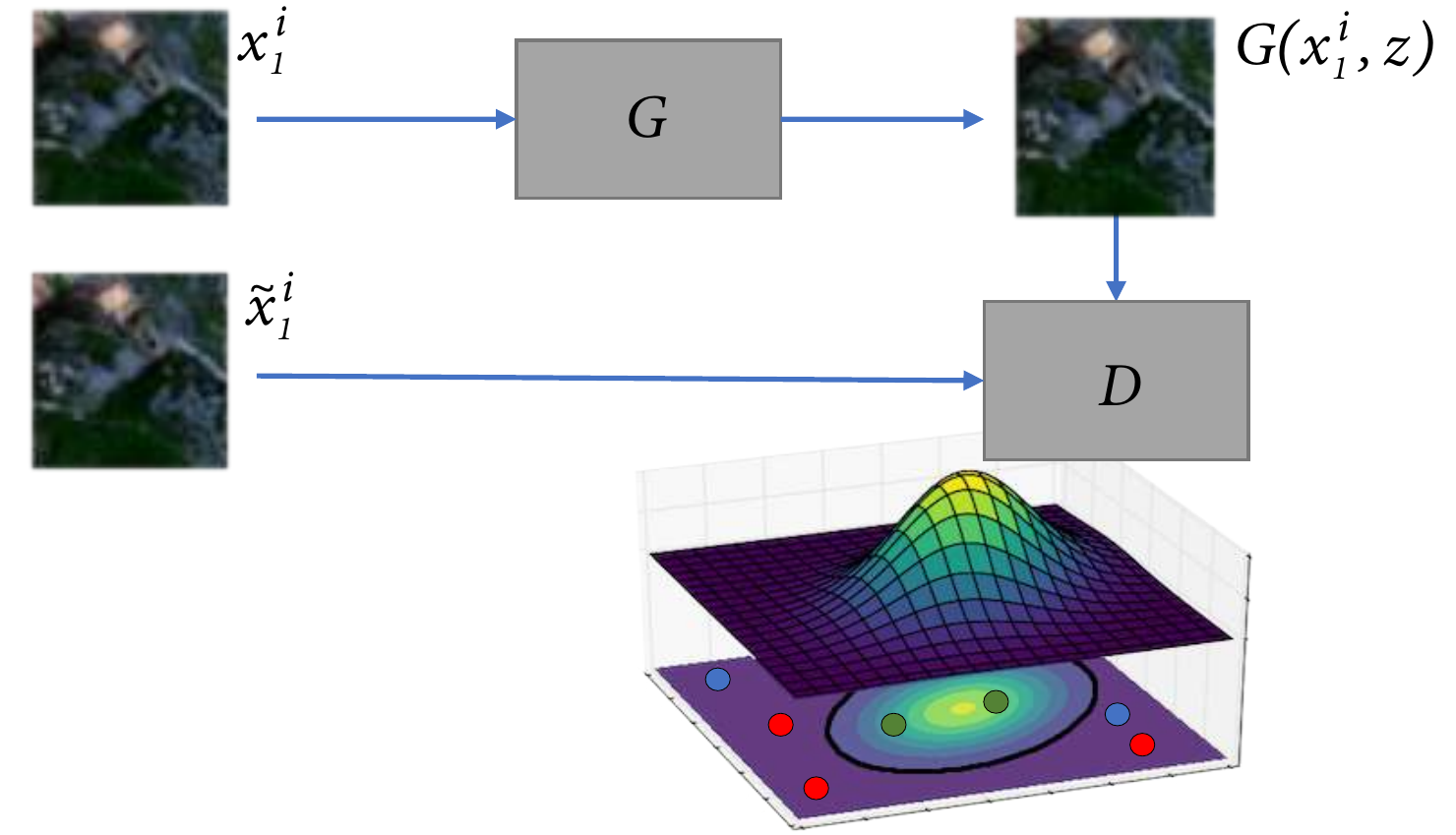}}
\caption{Illustration of the training strategy introduced within the proposed S\textsuperscript{2}-cGAN. The data distribution in the feature space (represented by the Gaussian curve under $D$) is much denser in the area corresponding to the real and unchanged samples. } 
\label{fig:teaser}
\end{figure}
Networks $G$ and $D$ are trained using both reconstruction and conditional GAN objectives. Reconstruction objective ${\cal L}_{L1}$ is given as:

\begin{equation} \label{eq:l1}
{\cal L}_{L1}(x^i_1,\tilde{x}_2^i) =  ||\tilde{x}_2^i - r ||_1,
\end{equation}

\noindent
where $r = G(x^i_1,z)$ and $z$ is a noise vector (drawn from a noise distribution ${\cal Z}$). The conditional adversarial objective ${\cal L}_{cGAN}$ is defined as:
\begin{align}
\begin{split}
{\cal L}_{cGAN}(G,D)= 
\mathbb{E}_{(x_1^i,\tilde{x}_2^i) \in {\cal X}} [\log D(x_1^i,\tilde{x}_2^i)] + \\
\mathbb{E}_{x_1^i \in  \boldsymbol{X_1} , z \in {\cal Z}} [\log ( 1 - D(x_1^i,G(x_1^i,z)) )],
\end{split}
\end{align}

It is important to emphasise that both images $ \boldsymbol{X_1}$ and $\boldsymbol{\tilde{X}}_2$ are unchanged co-registered images. The proposed learning approach does not need samples showing any change at training time. This makes it possible to train the discriminator without any need for fully supervised training data: $G$ acts as implicit supervision for $D$. During training, $G$ observes only unchanged patches. On the other hand, $D$ during training learns to distinguish real unchanged samples from those of changed or fake. As a consequence, at the end of the training process, the discriminator learns to separate real samples from artifacts.
The training procedure is schematically represented in Fig.~\ref{fig:teaser}. The data distribution is depicted by the Gaussian curve on the top of the figure. The discriminator is represented by the decision boundary on the learned feature space (black circle), which separates these distributions from the rest of the feature space. Both non-realistic generated samples (blue dots), and changed samples (red dots) are placed outside this decision boundary. The latter represents a case that it never observed during the training phase and hence is treated by $D$ as outliers (lies outside the discriminator's decision boundaries). The learned decision boundary of the discriminator is used to detect changes.

\begin{figure}[t]
\centerline{\includegraphics[width=0.9\linewidth]{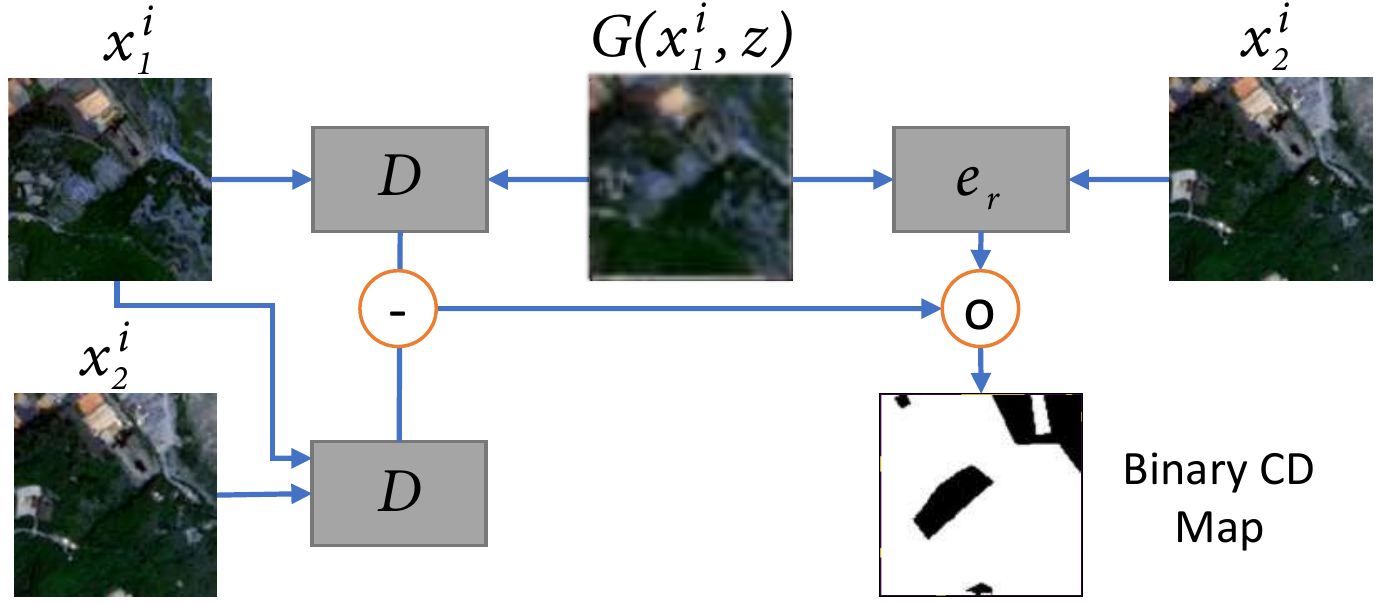}}
\caption{Illustration of the change detection strategy introduced within the proposed S\textsuperscript{2}-cGAN.}
\label{fig:cd}
\end{figure}

\subsection{CD through Adversarially Learned Representations}
\label{sec:cd}
In our adversarial approach for detecting changes, the discriminator network is used. As depicted in Fig. \ref{fig:cd}, given a test patch $x_1^i \in \boldsymbol X_1$, for time $t_1$ and its corresponding patch $x_2^i \in \boldsymbol X_2$ from time $t_2$. Given patch $x_1^i$, we utilize $G$ to generate $G(x_1^i)$ and compute the reconstruction error using (\ref{eq:l1}) with respect to $x_2^i$, where $e_r = {\cal L}_{L1}(x^i_2,G(x_1^i))$. In detail, we use the reconstruction errors of our adversarially trained generators as the first component of our CD strategy. In addition, we apply the pixel-based discriminators $D$ to estimate the out-of-distribution likelihood as our second component of CD. For that two pixel-wise score maps $S^{x_1^i,x_2^i}$ and $S^{x_1^i,G(x_1^i,z)}$ are computed using $D(x_1^i,x_2^i)$ and $D(x_1^i,G(x_1^i,z))$, respectively. 
Finally, a difference map $S_{dif}^i = S^{x_1^i,x_2^i}-S^{x_1^i,G(x_1^i,z)}$ is computed to represent the significance of score map $S^{x_1^i,x_2^i}$ considering the reference score map $S^{x_1^i,G(x_1^i,z)}$. 
As a result, a possible change between in $x_1^i$ and $x_2^i$, and/or between $G(x_1^i)$ and $x_2^i$ correspond to an outlier with respect to the data distribution learned by $D$ during training. This results in a low value in $S_{dif}$. The final CD map is obtained by Hadamard product for each pair of $e_r$ and $S_{dif}$, $\mathcal{H} = e_r \circ S_{dif}$.


\begin{figure}[t]

\begin{minipage}[b]{1.0\linewidth}
  \centering
  \frame{\includegraphics[width=.18\textwidth]{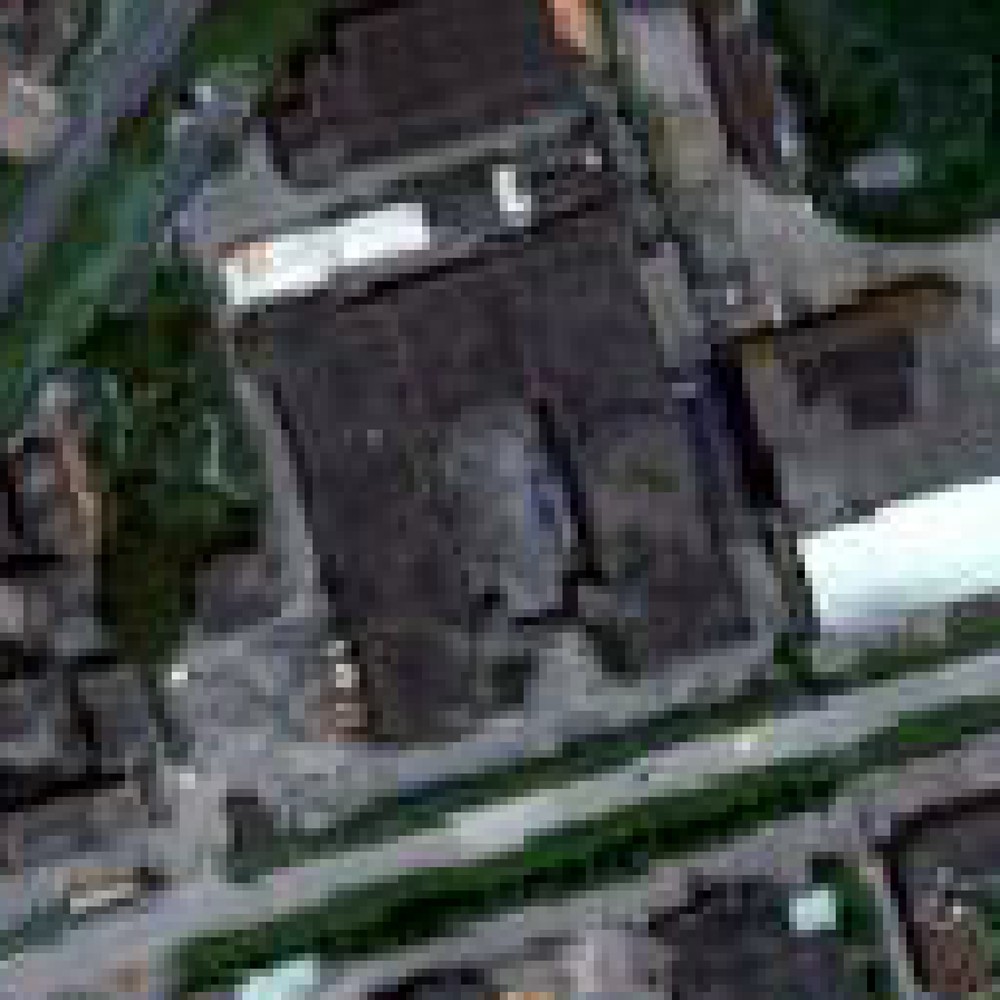}}
  \frame{\includegraphics[width=.18\textwidth]{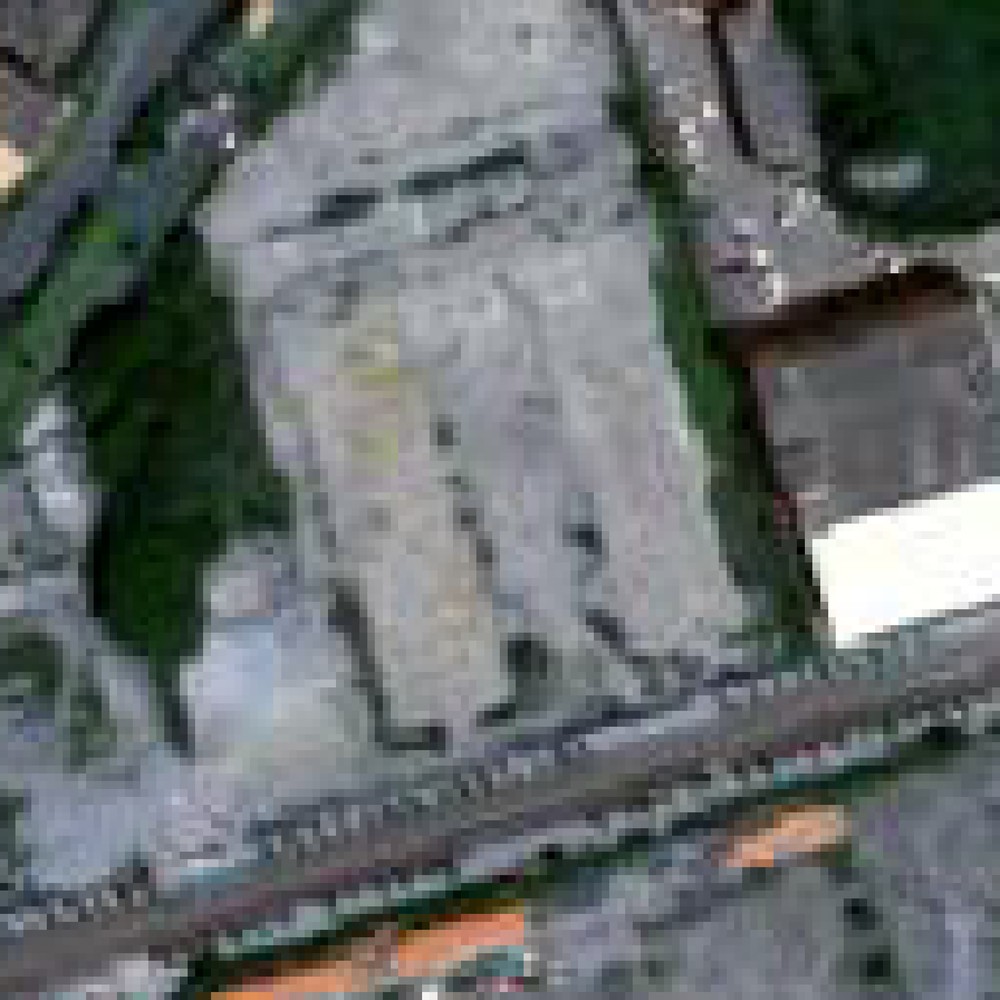}}
  \frame{\includegraphics[width=.18\textwidth]{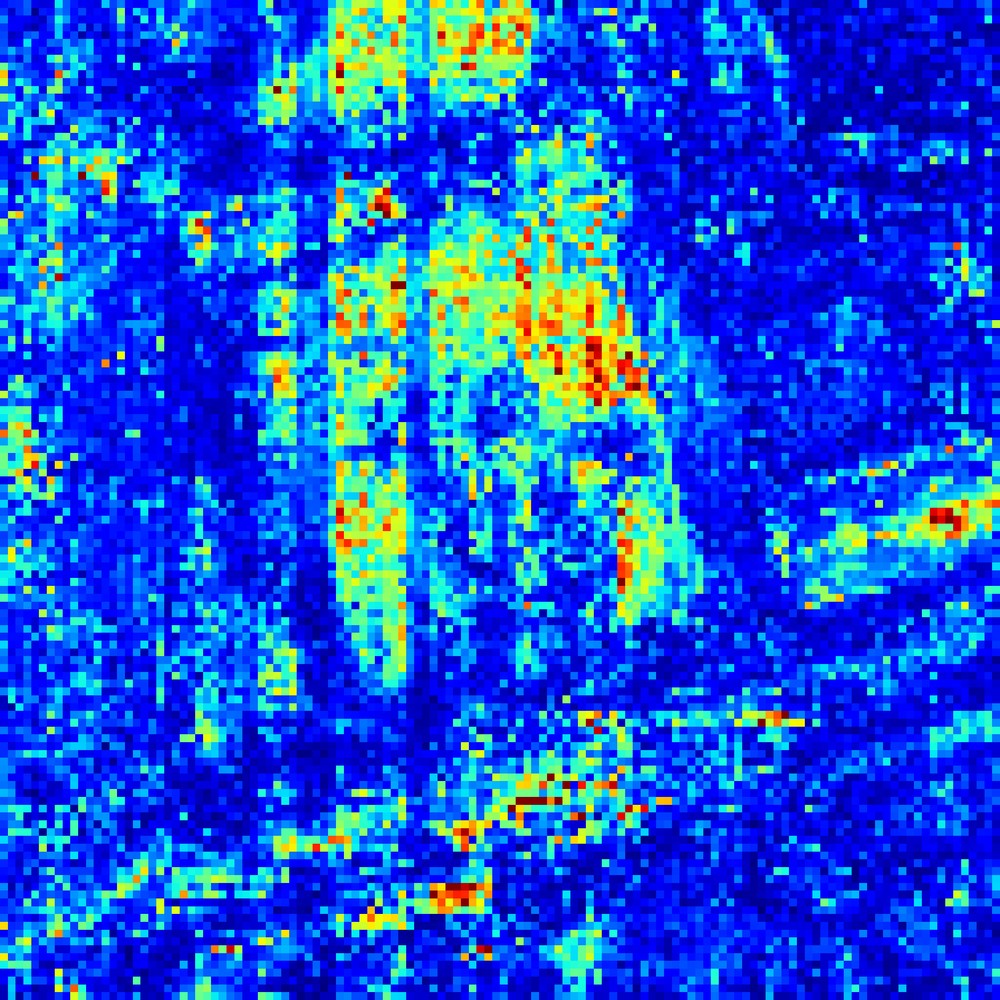}}
  \frame{\includegraphics[width=.18\textwidth]{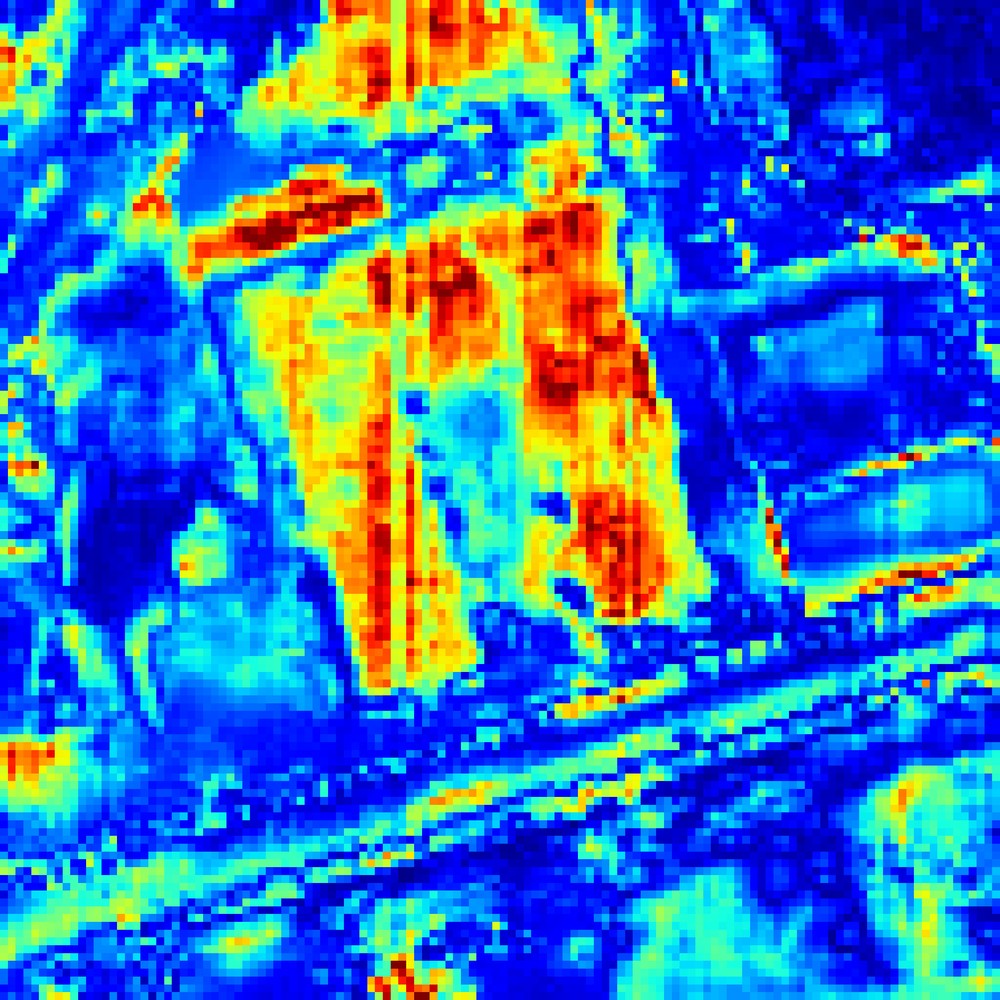}}
\frame{\includegraphics[width=.18\textwidth]{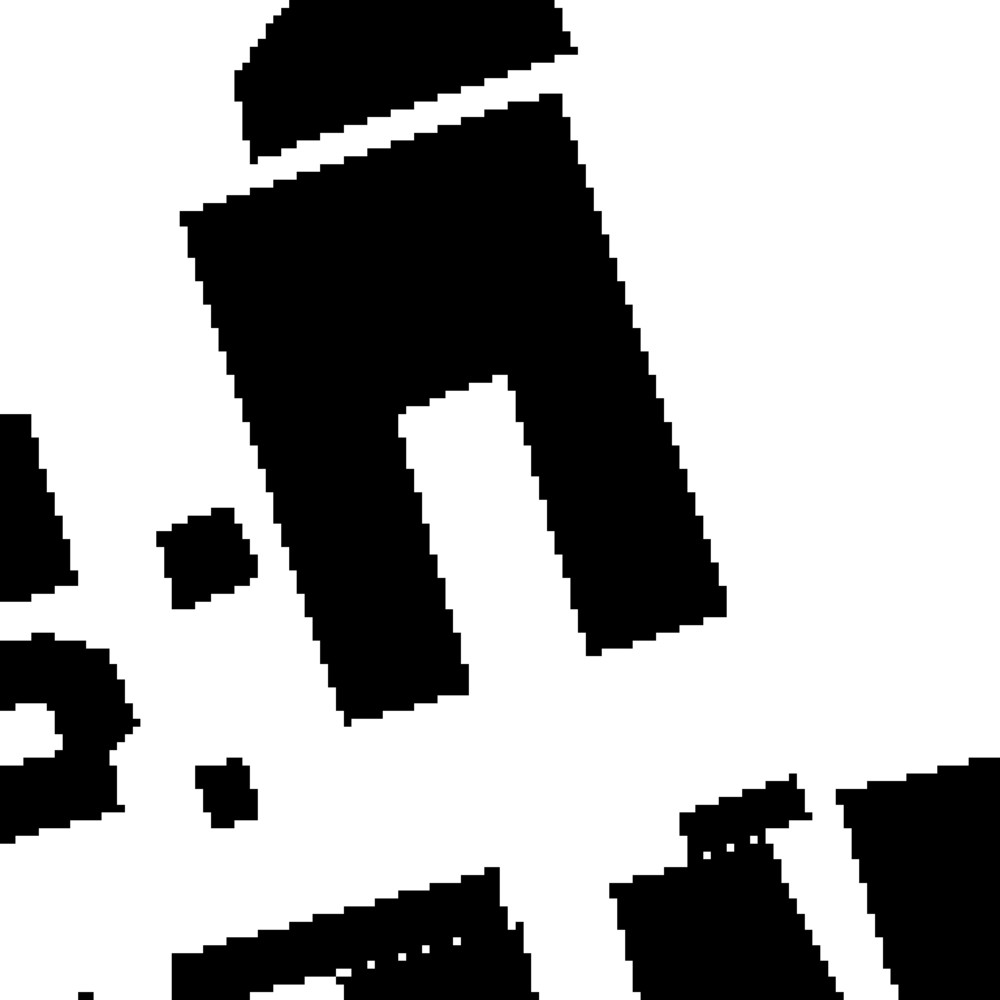}}
\vspace{.1cm}
\end{minipage}
\begin{minipage}[b]{1.0\linewidth}
  \centering
  \frame{\includegraphics[width=.18\textwidth]{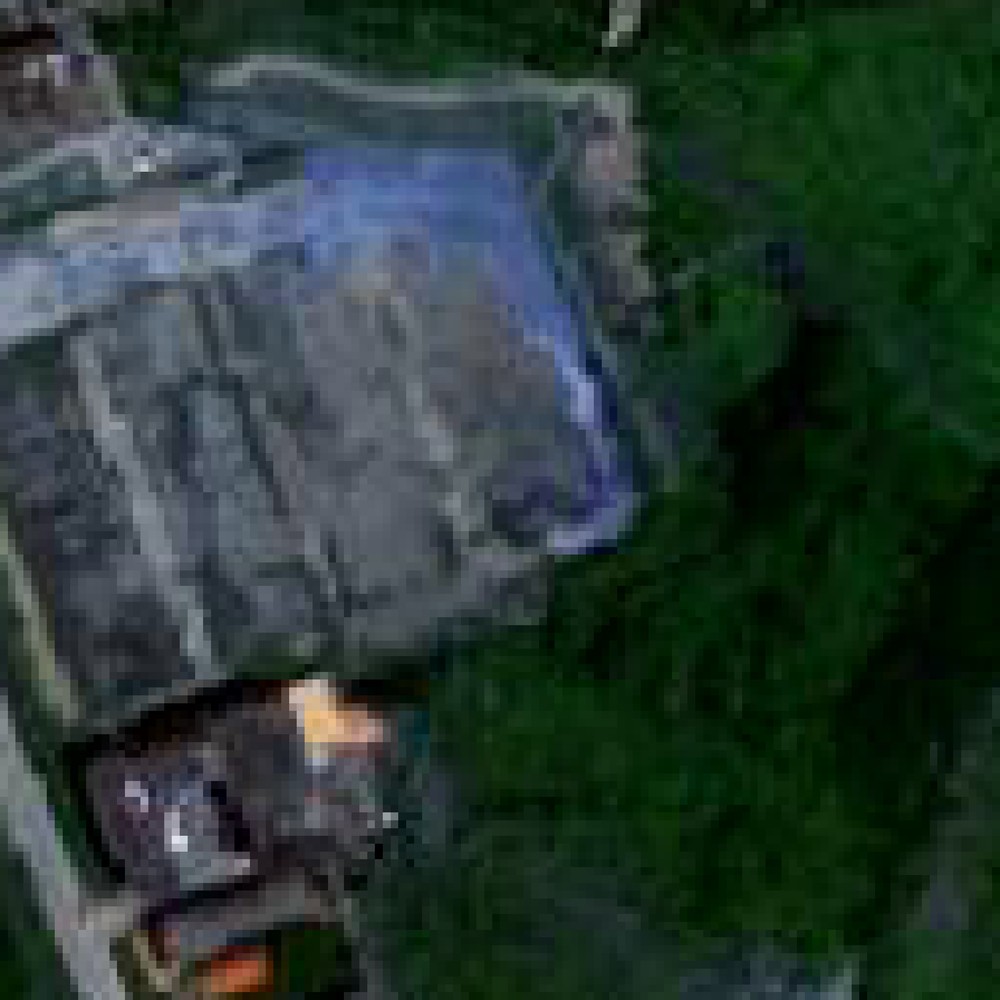}}
  \frame{\includegraphics[width=.18\textwidth]{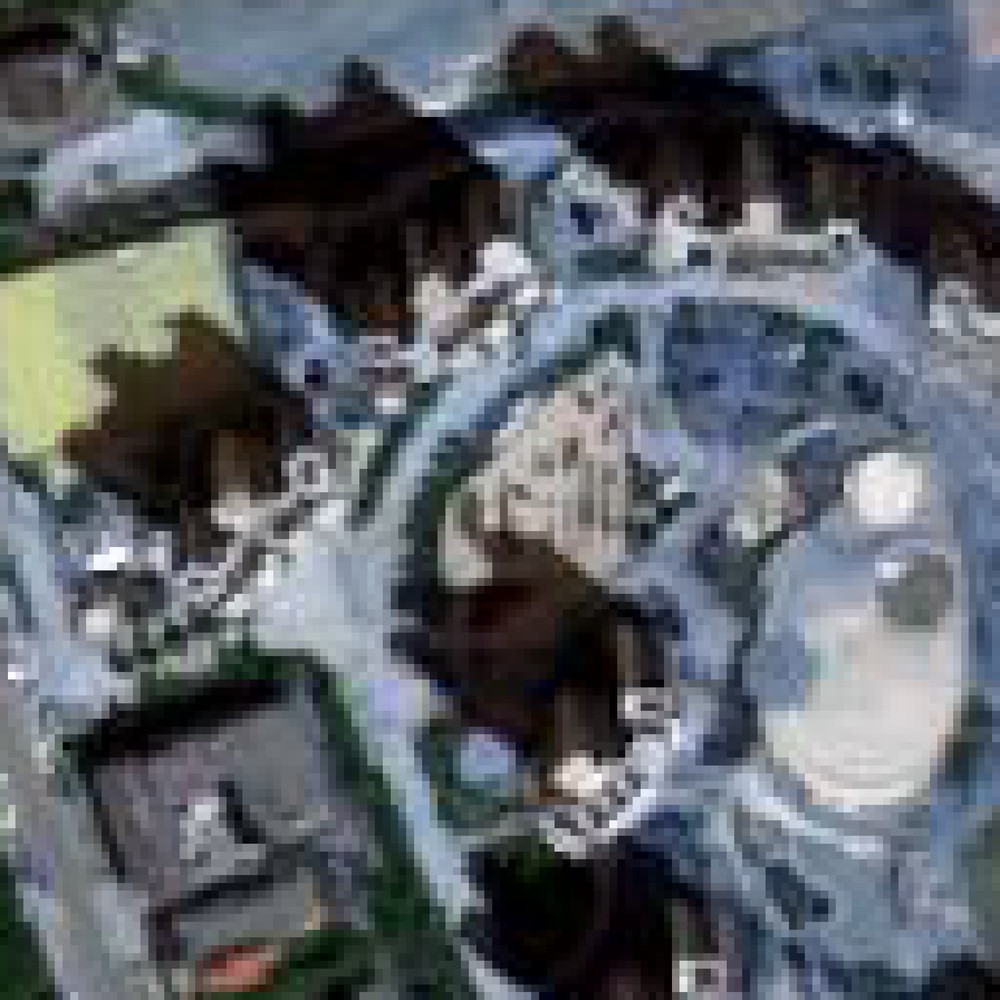}}
  \frame{\includegraphics[width=.18\textwidth]{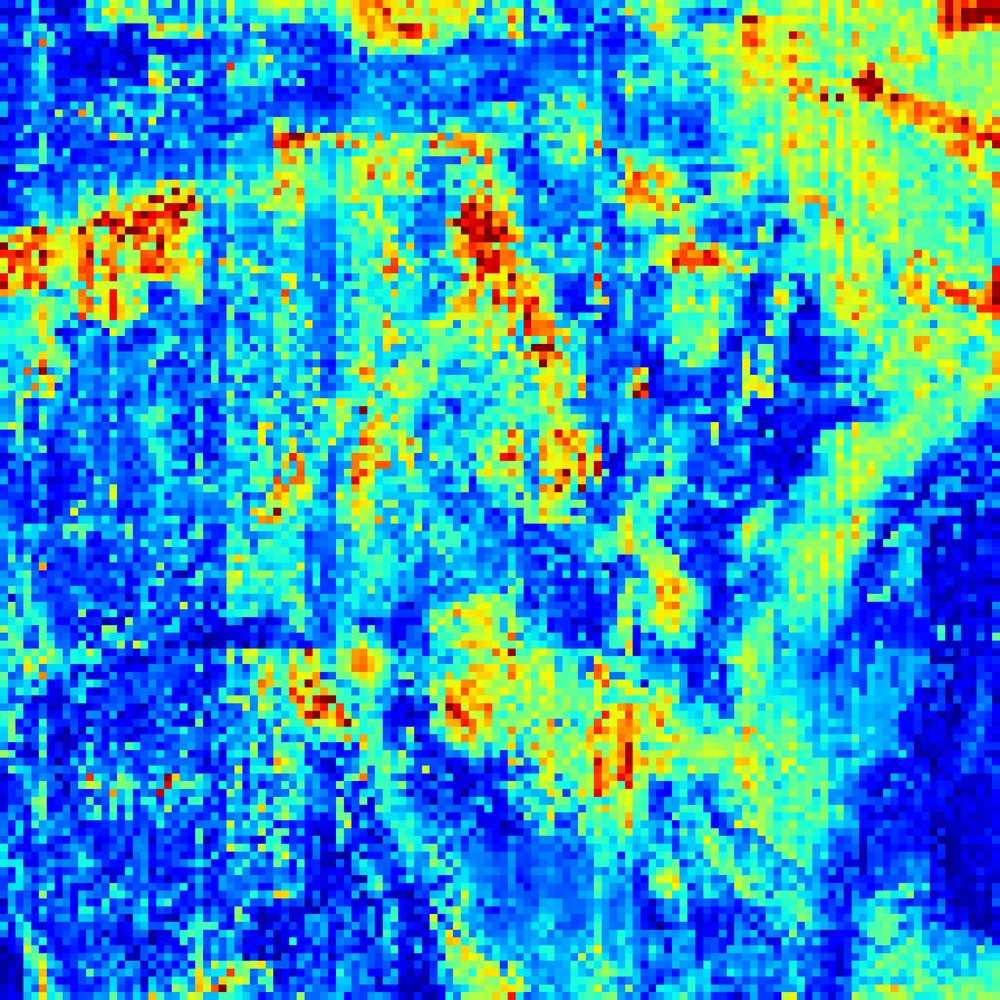}}
  \frame{\includegraphics[width=.18\textwidth]{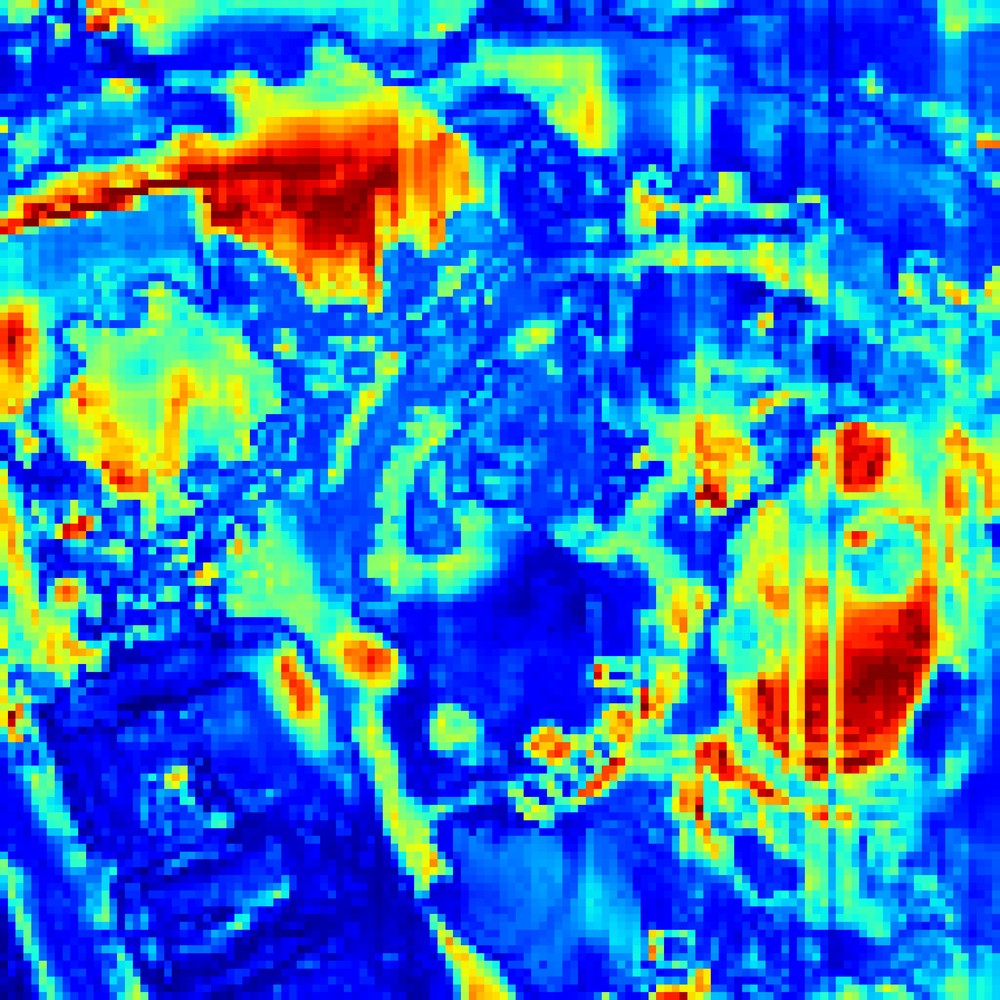}}
\frame{\includegraphics[width=.18\textwidth]{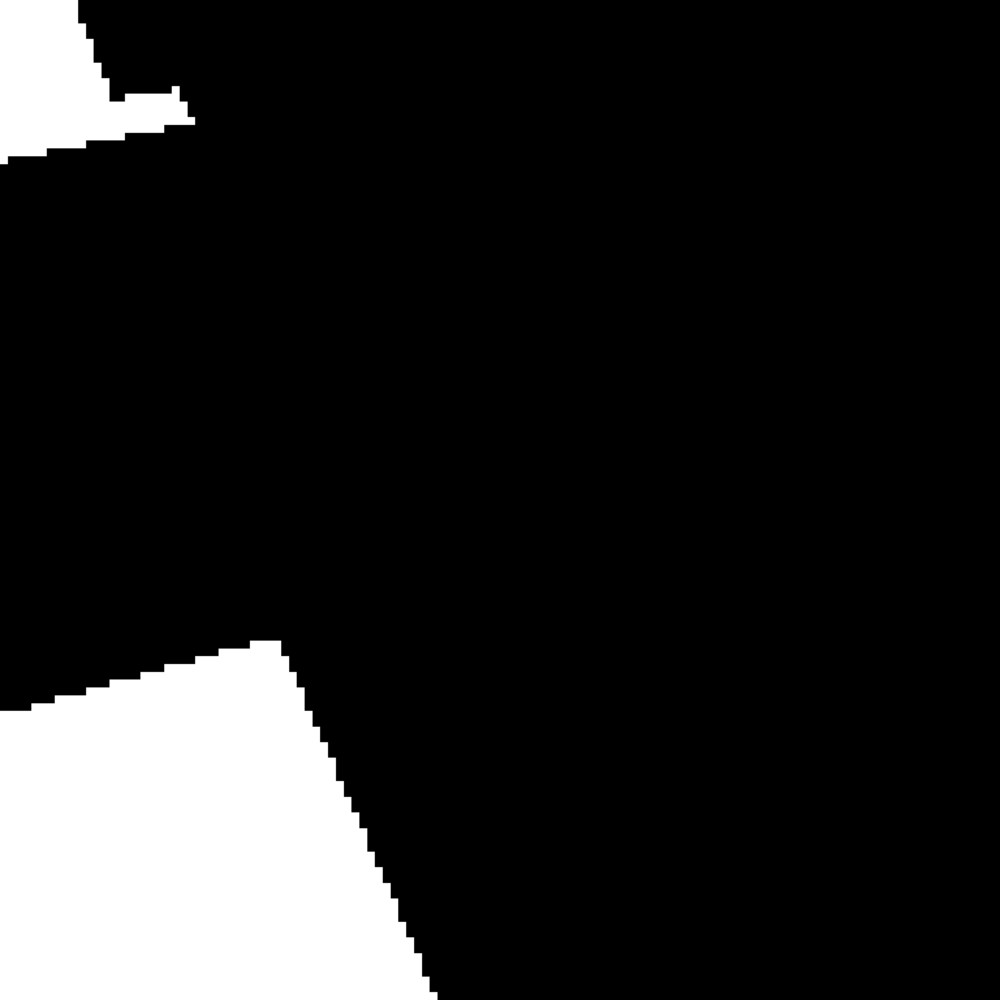}}
\vspace{.1cm}
\end{minipage}

\begin{minipage}[b]{1.0\linewidth}
  \centering
\setlength\tabcolsep{18pt}
\begin{tabular}{ccccc}
\centering
      (a)  & (b) &  (c)  &  (d)  &  (e)  \\
\end{tabular}
\end{minipage}
\caption{Examples of score maps associated to different pairs of patches. For each row: (a) a patch acquired at $t_1$ ($x_1^i$); (b) a patch acquired at $t_2$ ($x_2^i$); (c) their differences map ($S_{dif}$); (d) their reconstruction error ($e_r$); and (e) reference CD map.}

\label{fig:res}
  \vspace{-0.2cm}
\end{figure}

\section{Experimental Results}
\label{sec:exp}

\begin{figure}[htb]

\begin{minipage}[b]{1\linewidth}
  \centering
\setlength\tabcolsep{8pt}
\begin{tabular}{cc}
\centering
  \frame{\includegraphics[width=.35\textwidth]{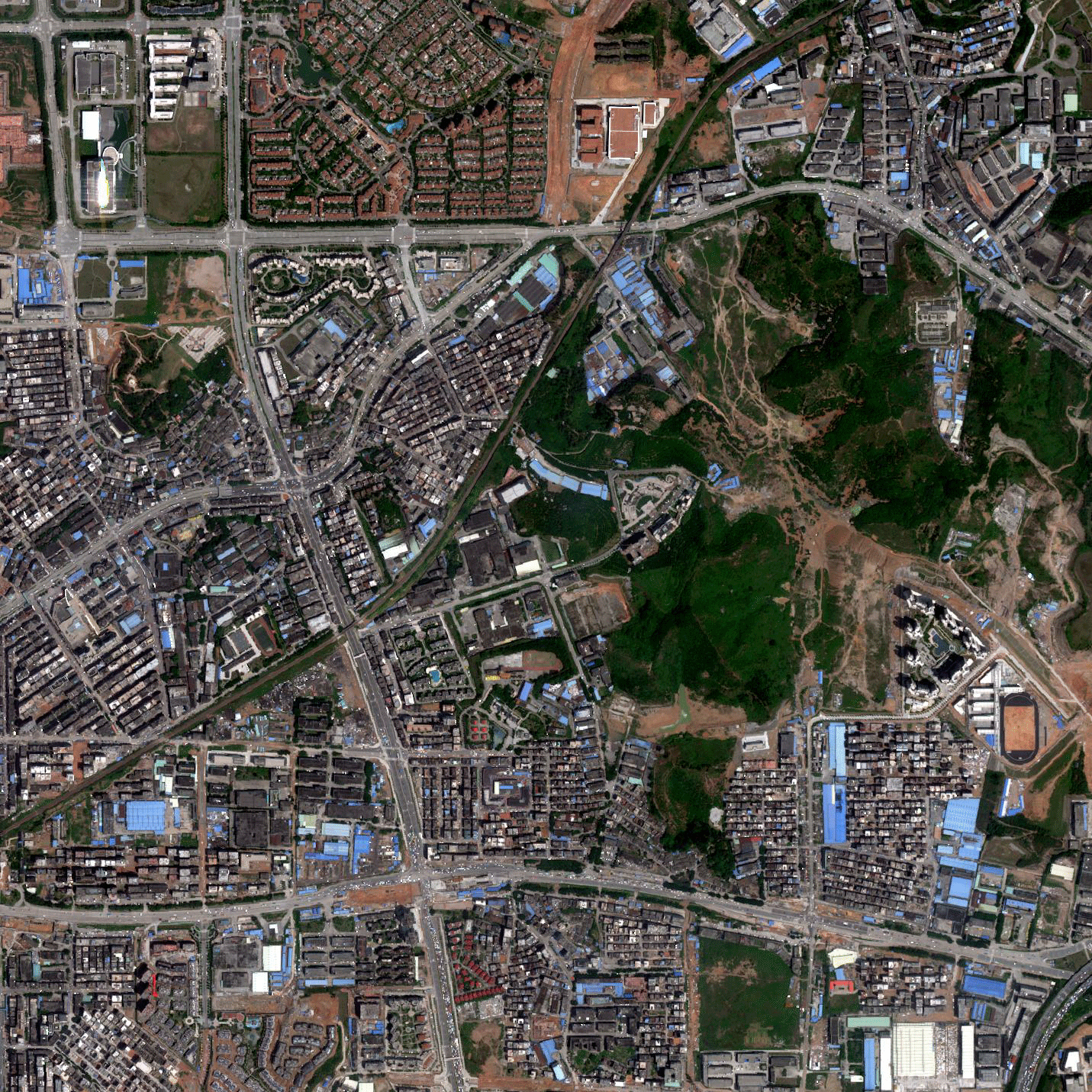}}&
  \frame{\includegraphics[width=.35\textwidth]{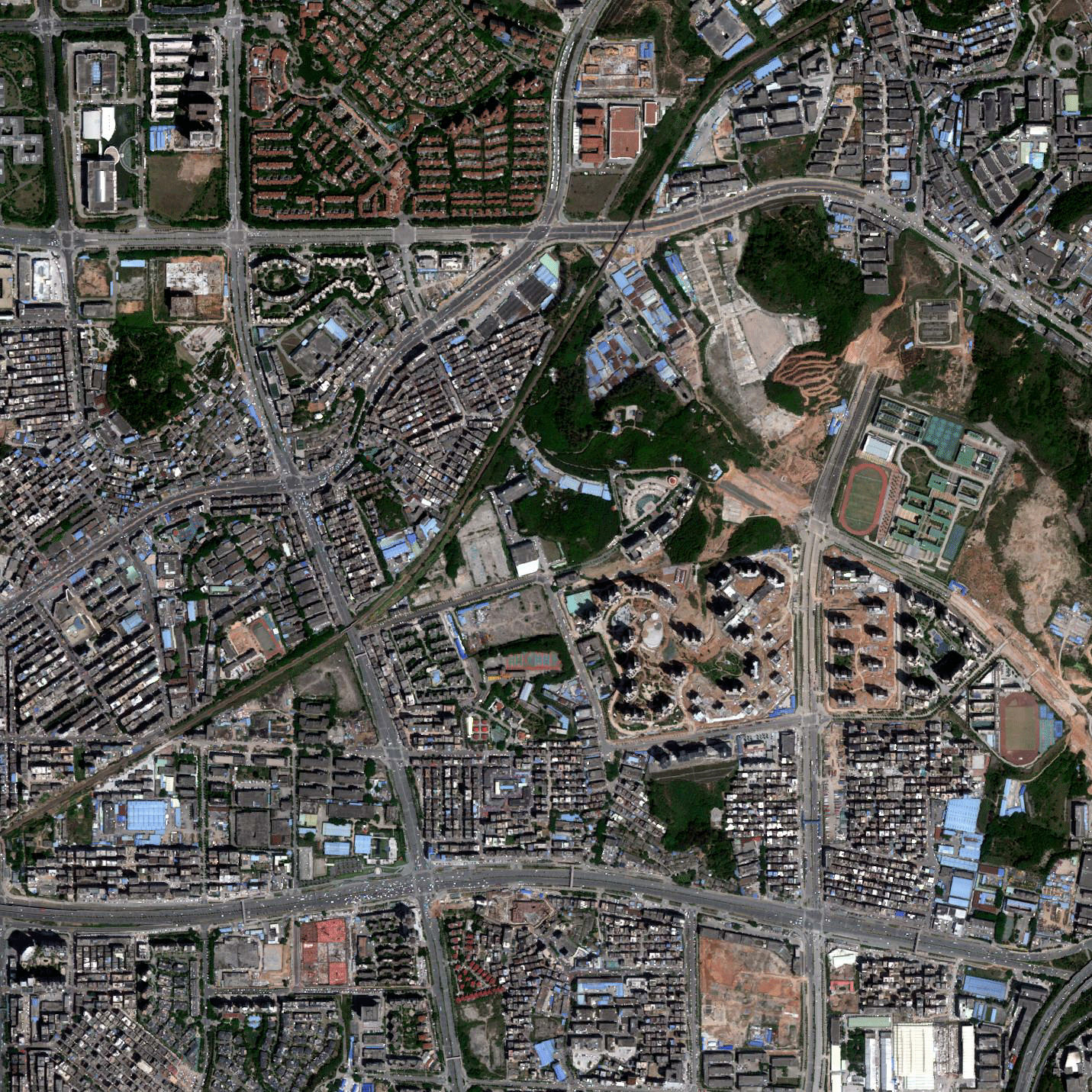}}\\
  (a) & (b) \\ 
  \frame{\includegraphics[width=.35\textwidth]{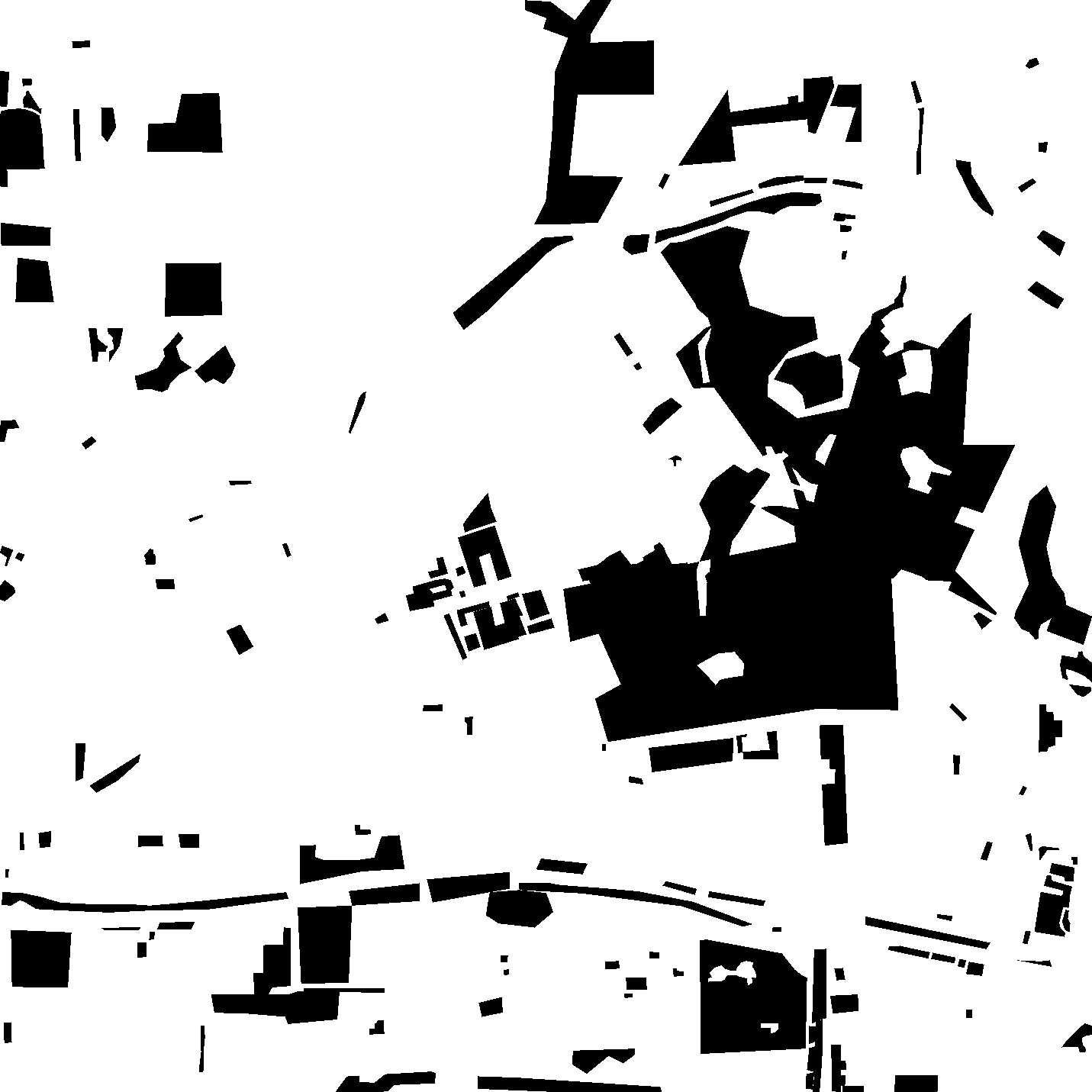}}&
  \frame{\includegraphics[width=.35\textwidth]{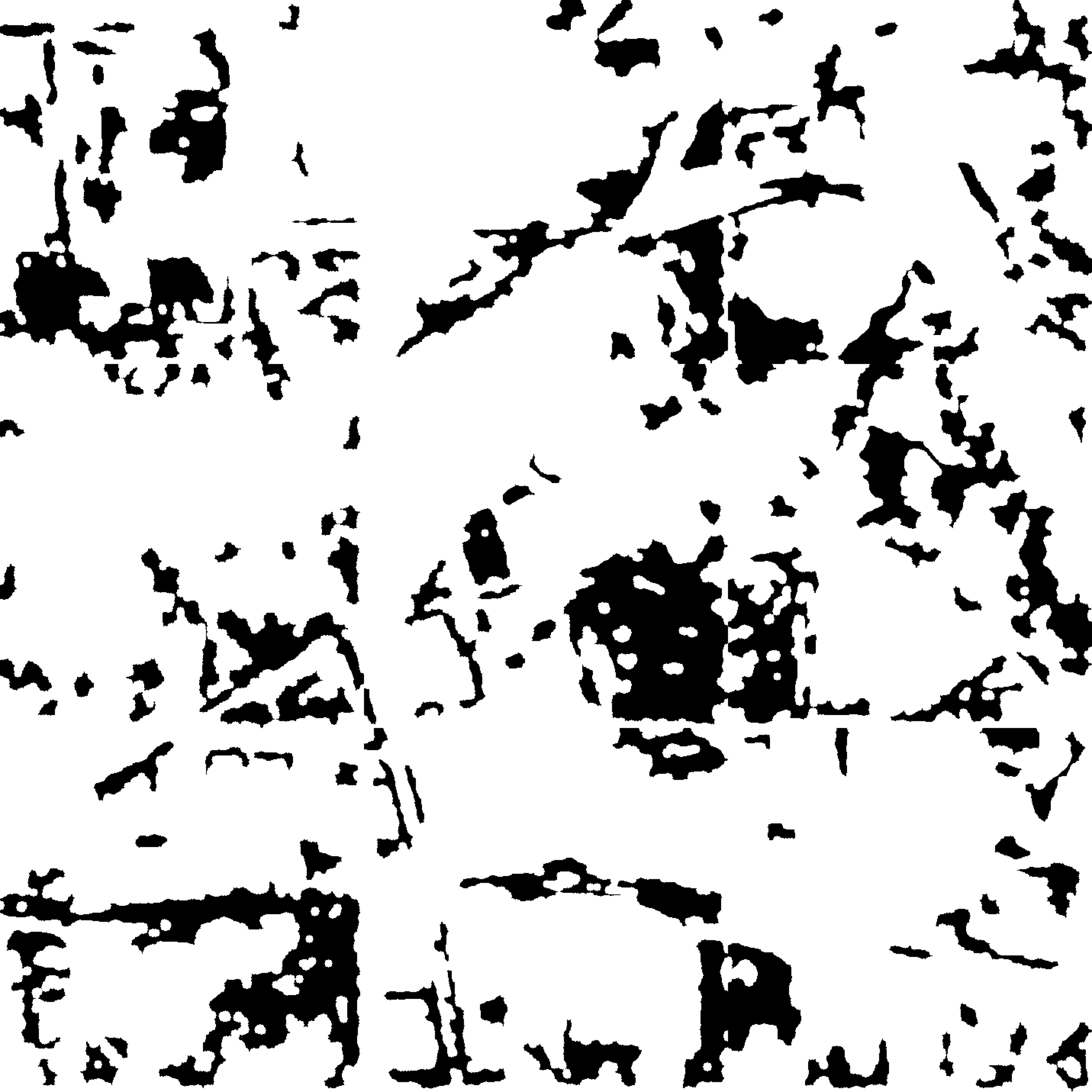}}\\
  (c) &  (d)  \\ 
  \frame{\includegraphics[width=.35\textwidth]{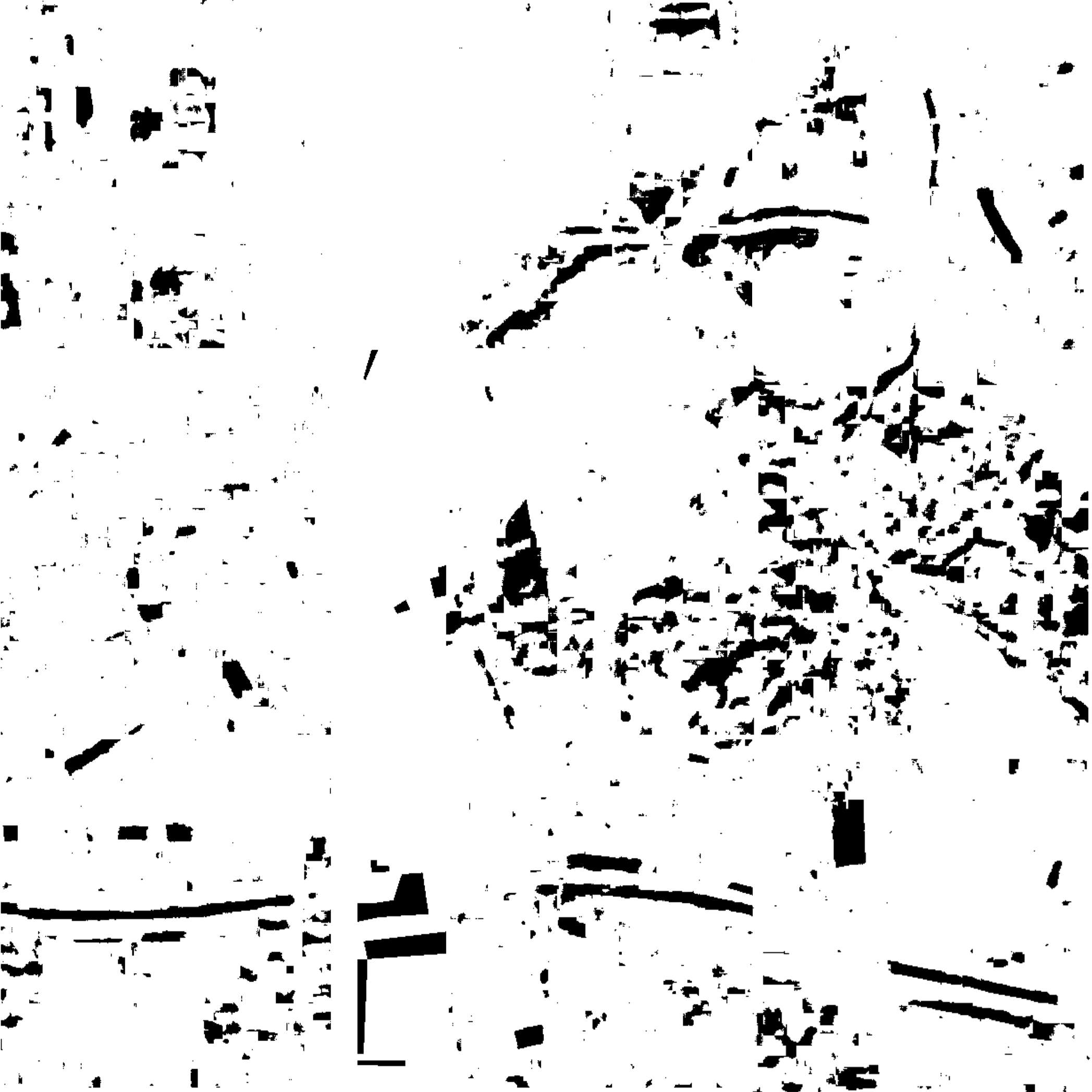}}&
\frame{\includegraphics[width=.35\textwidth]{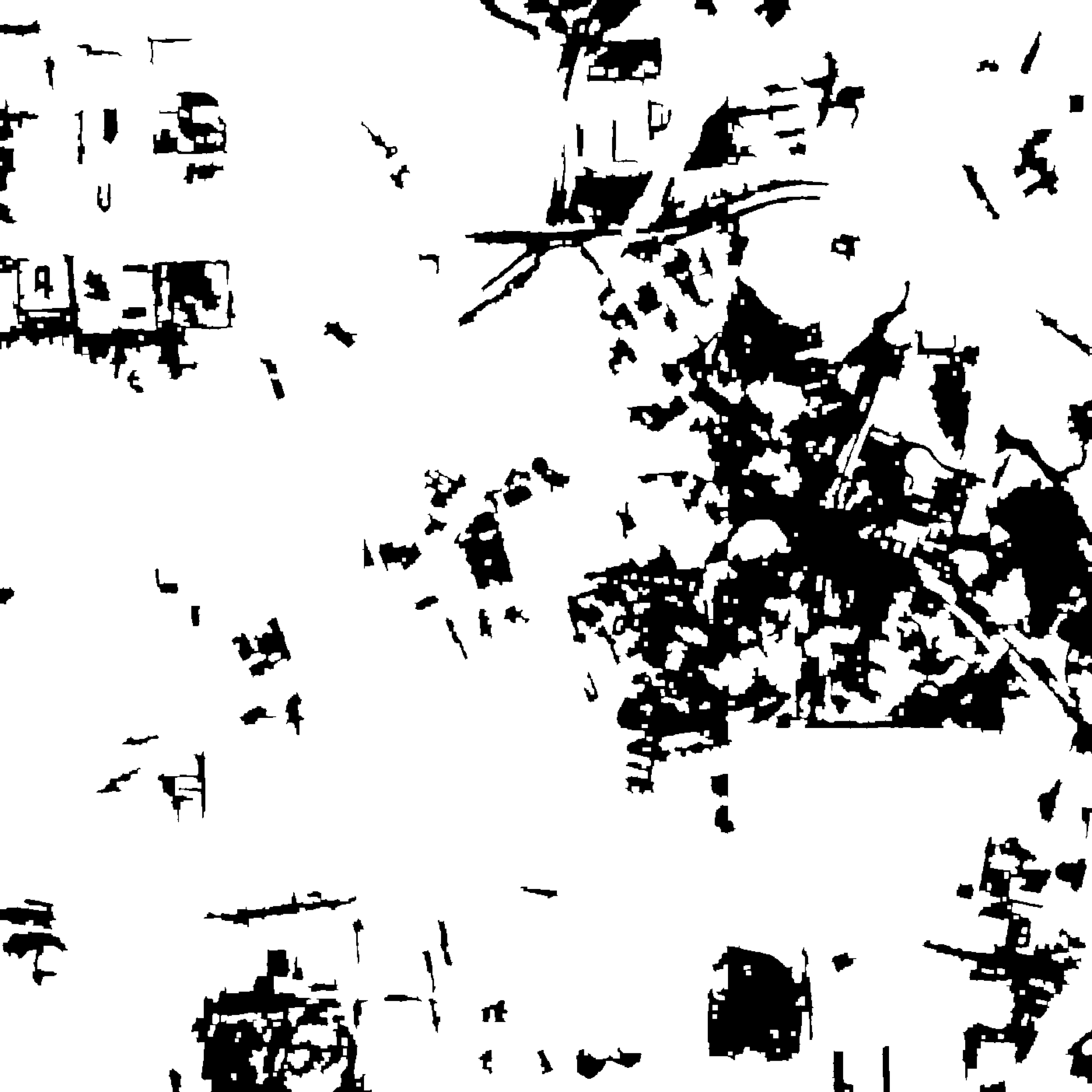}}\\
        (e) & (f) \\
\end{tabular}
\end{minipage}
\caption{CD results: (a)  $\boldsymbol X_1$; (b) $\boldsymbol X_2$; (c) reference CD map; (d) CD map obtained by the DCVA; (e) CD map obtained by the FC-EF; and (f) CD map obtained by the S\textsuperscript{2}-cGAN.}

\label{fig:res2}
  \vspace{-0.2cm}
\end{figure}

To evaluate the proposed S\textsuperscript{2}-cGAN, in the experiments a bitemporal Very High Spatial Resolution (VHR) multispectral images acquired from Shenzhen, China by the Worldview 2 satellite has been used \cite{9052762}. 
The image acquired in 2010 ($t_1$) has been considered as $X_1$, while the image acquired in 2015 ($t_2$) has been taken as $X_2$. Both $X_2$ and $X_2$ have size of $1431\times1431$ pixels with a spatial resolution of 2 m.
In all the experiments, the generator network $G$ was based on a U-net autoencoder for input patches of size $128\times128$ pixels. The generator was compound by a network with six convolutional layers. For discriminator $D$, a pixel-based network with two convolutional layers was employed. The co-registered images $\boldsymbol X_1$ and $\boldsymbol X_2$ were divided into 7646 number of patches of size $128 \times 128$ pixels. For the self-supervised adversarial learning of the unchanged sample representations, we construct  $\boldsymbol{\tilde X}_2$ from $\boldsymbol X_1$ by applying a Gaussian noise over $\boldsymbol X_1$. This represents the unchanged (generated) image at time $t_2$. Half of the pairs of patches of $\boldsymbol X_1$ and $\boldsymbol{\tilde X}_2$ were randomly selected for training. The remaining patches of $\boldsymbol X_1$, together with their pairs in $\boldsymbol X_2$ were considered as pairs of test patches for the evaluation. The training was based on stochastic gradient descent with momentum 0.5, and the network was trained for 50 epochs. The two score maps $S_f$ (the differences map) and $e_r$ (the reconstruction error) were computed as explained in Sec. \ref{sec:cd}. 
Fig. \ref{fig:res} illustrates the score maps for two different pairs of patches. From the figure, one can observe that in most of the cases $e_r$ and $S_f$ provide complementary information. Similar behaviour has been observed by varying the pairs of patches. For quantitative evaluation, we applied a threshold value $\tau$ over the score map $\mathcal{H}$ to obtain the binary CD map. Due to the strong local variations in the considered VHR images, instead of choosing a single decision boundary for $\tau$ value, we adapted the context-dependent local adaptive decision boundary strategy used in \cite{saha2018unsupervised}. 
We compared the CD map obtained by the proposed S\textsuperscript{2}-cGAN with: 1) Fully Convolutional Early Fusion (FC-EF) method, which is one of the best performing supervised methods with a fully convolutional U-Net architecture \cite{daudt2018fully}; and 2) Deep Change Vector Analysis (DCVA) method, which is a powerful unsupervised method using a CNN pre-trained for semantic segmentation to obtain multi-temporal deep features \cite{saha2018unsupervised}. For the sake of fairness, the same pairs of test patches have been used for all the considered methods. Note that DCVA is fully unsupervised and does not use labeled samples. The fully-supervised FC-EF exploits a set of pairs of training patches that consists of: i) the same pairs of labeled unchanged samples used in the S\textsuperscript{2}-cGAN; and ii) 13,065 pairs of changed samples. The binary CD maps obtained by the FC-EF, DCVA and the proposed S\textsuperscript{2}-cGAN are shown in Fig. \ref{fig:res2}. From Fig. \ref{fig:res2}, one can see that the proposed S\textsuperscript{2}-cGAN in most cases is able to identify the central area of the changed objects correctly. Table \ref{tab:results} shows quantitative results in terms of overall accuracy (OA), specificity (SPC), sensitivity (SEN), and overall error rate (ERR) obtained by the FC-EF, the DCVA, and the proposed S\textsuperscript{2}-cGAN. By analysing the table, one can observe that the proposed S\textsuperscript{2}-cGAN is comparable with the fully supervised FC-EF, and outperforms the unsupervised method DCVA mainly in terms of specificity (which shows the accuracy of detecting unchanged samples), sensitivity (which shows the accuracy of detecting changed pixels) and overall accuracy. As an example, sensitivity measure obtained by the proposed method is 4\% and 14\% higher than those obtained by the unsupervised DCVA and fully supervised FC-EF, respectively. It is worth emphasizing that our method performs better than both methods in terms of sensitivity despite the absence of changed samples during the training. This is due to the impact of adversarial training during GAN optimization. 
\begin{table}[t]
\centering
\caption{\small{Binary CD results obtained by the FC-EF, the DCVA, and the proposed S\textsuperscript{2}-cGAN. OA: overall accuracy, SPC: Specificity, SEN: Sensitivity, and ERR: Overall Error Rate.}}
\scalebox{0.83}{

{\renewcommand{\arraystretch}{1.3}
\begin{tabular}{llcccc}\hline
 Method    &  Category           &OA     & SPC   & SEN   & ERR \\ \hline
    FC-EF &Supervised    &0.8633 &0.9693 &0.3627 &0.1366 \\ 
    DCVA &Unsupervised   &0.8280 &0.9106 &0.4450 &0.1719\\ 
   \textbf{S\textsuperscript{2}-cGAN}&Self-Supervised     &0.8482 &0.92081 &0.5053 &0.1517\\ \hline
\end{tabular}}
}
\label{tab:results}
\vspace{-0.3cm}
\end{table}

\vspace{-0.3cm}
\section{Conclusion}
\label{sec:subhead}

In this paper, we have introduced a self-supervised conditional Generative Adversarial Network (S\textsuperscript{2}-cGAN) for binary CD problems in RS. The proposed S\textsuperscript{2}-cGAN exploits the mutual supervisory information of the generator and the discriminator networks to train a deep network by using a self-supervised multitemporal training set (which includes only pairs of unchanged samples). Differently from the existing GAN based CD methods, the proposed method directly uses the GAN discriminator as the classifier. Experimental results show that the proposed S\textsuperscript{2}-cGAN leads to a higher performance in terms of sensitivity compared to the state of the art fully supervised and unsupervised methods. This has been achieved without using any pairs of labeled change samples. We underline that this is a very important advantage, since the proposed S\textsuperscript{2}-cGAN suppresses the cost required for reference data collection. As a future work, we plan to extend the proposed learning approach to be integrated in an active learning setup with the human in the loop. 

\vspace{-0.1cm}
\section{Acknowledgement}
\vspace{-0.1cm}
This work was supported by the European Research Council under the ERC Starting Grant BigEarth-759764.
\vspace{-0.05cm}



\bibliographystyle{IEEEbib}
\small{
\bibliography{refs}
}

\end{document}